\documentclass[%
 reprint,
 superscriptaddress,
 amsmath,
 aps, prl,
]{revtex4-1}

\usepackage{graphicx}
\usepackage{dcolumn}
\usepackage{bm}
\usepackage{xcolor}
\usepackage{ulem}

\definecolor{mygreen}{rgb}{0., 0.5, 0.0}

\usepackage{soul}

\begin{document}

\preprint{APS/123-QED}

\title{Thermal evolution of spin excitations in honeycomb Ising antiferromagnetic FePSe$_3$}

\author{Lebing Chen}
 \email{lebingchen@berkeley.edu}
\affiliation{Department of Physics and Astronomy, Rice University, Houston, Texas, 77005, USA}
\affiliation{Department of Physics, University of California at Berkeley, Berkeley, California, 94720, USA}
\affiliation{Material Sciences Division, Lawrence Berkeley National Lab, Berkeley, California, 94720, USA}

\author{Xiaokun Teng}
\affiliation{Department of Physics and Astronomy, Rice University, Houston, Texas, 77005, USA}
\author{Ding Hu}
\affiliation{School of Physics, Hangzhou Normal University, Hangzhou, Zhejiang, 311121, People's Republic of China}
\author{Feng Ye}
\author{Garrett E. Granroth}
\affiliation{Neutron Scattering Division, Oak Ridge National Laboratory, Oak Ridge, Tennessee, 37831, USA}
\author{Ming Yi}
\affiliation{Department of Physics and Astronomy, Rice University, Houston, Texas, 77005, USA}
\author{Jae-Ho Chung}
 \email{jaehc@korea.ac.kr}
\affiliation{Department of Physics, Korea University, Seoul 02841, Korea}
\author{Robert J. Birgeneau}
 \email{robertjb@berkeley.edu}
\affiliation{Department of Physics, University of California at Berkeley, Berkeley, California, 94720, USA}
\affiliation{Material Sciences Division, Lawrence Berkeley National Lab, Berkeley, California, 94720, USA}
\author{Pengcheng Dai}
 \email{pdai@rice.edu}
\affiliation{Department of Physics and Astronomy, Rice University, Houston, Texas, 77005, USA}

\date{\today}

\begin{abstract}
We use elastic and inelastic neutron scattering (INS) to study the antiferromagnetic (AF) phase transitions and spin excitations in the two-dimensional (2D) zig-zag antiferromagnet FePSe$_3$. By determining the magnetic order parameter across the AF phase transition, we conclude that the AF phase 
transition in FePSe$_3$ is first-order in nature. In addition, our INS measurements reveal that the spin waves in the AF ordered state have a large easy-axis magnetic anisotropy gap, consistent with an Ising Hamiltonian, and possible biquadratic magnetic exchange interactions. 
On warming across $T_N$, we find that dispersive spin excitations 
associated with three-fold rotational symmetric AF fluctuations change into FM spin fluctuations above $T_N$. These results suggest that  
the first-order AF phase transition in FePSe$_3$ may arise from the competition between 
$C_3$ symmetric AF and $C_1$ symmetric FM spin fluctuations around $T_N$, in place of a conventional second-order AF phase transition.  
\end{abstract}

\maketitle


Two-dimensional (2D) spin models are one of the most well-studied magnetic systems in condensed matter physics. The pioneering solution by Onsager in 1944 \cite{Onsager1944} revealed that the 2D ferromagnetic (FM) Ising model, characterized by nearest neighbor coupling $J$, undergoes a magnetic phase transition at $k_BT_C=2.269J$, exhibiting an order parameter critical exponent $\beta=1/8$. The 2D XY model, on the other hand, lacks long-range order, but it exhibits the ability to undergo a Berezinskii–Kosterlitz–Thouless transition, during which the correlation function transitions from an exponential decay to a power-law behavior as a function of distance\cite{B,KT}. In real magnetic materials, {spins exhibit free 3D rotation rather than adhering to binary orientations or 2D rotations}, and the Hamiltonian for actual magnetic systems will take the form of a Heisenberg model plus magnetic anisotropy \cite{Gibertini2019}. According to the Mermin-Wagner theorem, in the case of an isotropic Heisenberg Hamiltonian with short-range magnetic exchange couplings, the thermal fluctuations are strong enough to prevent long-range magnetic order in the 2D  limit at any finite temperature \cite{Mermin}, with the correlation length diverging only at absolute zero. Conversely, in the presence of easy-axis/easy-plane anisotropy, the renormalization group flow will favor the selection of the anisotropic component of the Hamiltonian as the relevant parameter, resulting in critical behavior that aligns with predictions from the Ising/XY models \cite{Goldenfeld1972}. 

The discovery of long-range magnetic order at non-zero temperatures in the 2D monolayer of 
several honeycomb lattice van der Waals (vdW) ferromagnets and antiferromagnets suggests a suppression of the thermal fluctuations, 
most likely due to the formation of an Ising-type magnetic anisotropy gap in these materials \cite{Huang2017,Gong2017,Lee2016,Lee2023,Luo2023}. To understand the microscopic origin of the long-range magnetic order in the 2D limit for different classes of vdW materials, it is therefore important to determine the magnetic properties of their 3D bulk
compounds.
While the FM Ising Hamiltonian on a 2D square lattice has been solved \cite{Onsager1944}, the situation for Ising spin systems in 2D honeycomb lattices is more complicated. As a non-Bravais lattice, the honeycomb structure can host a few different collinear magnetic structures,
including simple ferromagnets, N$\rm \acute{e}$el antiferromagnet with $c$-axis aligned moments, stripy antiferromagnet, and zig-zag antiferromagnet [Fig. 1(e)], 
depending on the relative strengths of nearest $J_1$ (NN), next nearest $J_2$ (NNN), and next-next nearest neighbor $J_3$ (NNNN) magnetic exchange interactions [Fig. 1(a)] \cite{Fouet2001}. The first two magnetic structures have the same in-plane ordering wavevectors as the lattice vectors, and respect the three-fold rotational ($C_3$) symmetry of the honeycomb lattice. The stripy and zig-zag antiferromagnetic (AF) structures, however, break the $C_3$ rotational symmetry of the honeycomb lattice to $C_1$ and fold the first Brillouin zone (FBZ) into a smaller rectangular magnetic FBZ. In these structures, there will be three magnetic domains separated by 60$^\circ$ degrees, meaning that the overall magnetism still obeys the underlying  $C_3$  lattice symmetry. 
The magnetic order parameter is not a direct measure of the ordered or staggered moment because it has multiple components. 
For example, the order parameter for the zig-zag AF order is composed of three components ($\Psi_1, \Psi_2, \Psi_3$), each reflecting the staggered moment along one of the three $C_3$ axes \cite{Domany1978}. By probing the static magnetic order and spin excitations across the
magnetic phase transition, one can obtain information concerning the behaviors of the phase transition and compare the results with the expectations of 
the conventional Ising or XY models.

For the FM honeycomb system CrSiTe$_3$, the critical behavior of the magnetic phase transition exhibits 
2D Ising characteristics \cite{Liu2016,Williams2015}, with a dimensional crossover from 2D to 3D near $T_C$\cite{Ron2019}. For other honeycomb lattice ferromagnets such as CrI$_3$, VI$_3$, and CrGeTe$_3$, the critical behavior is more reminiscent of the 
tricritical mean field model \cite{Liu2018,Chen2020,Liu2020,Hao2021,Lin2017}. For AF ordered MnPS$_3$, NiPS$_3$ and CoPS$_3$, the critical behavior near $T_N$ shows 3D XY or Ising characteristics \cite{Wildes2007,Wildes2017,Wildes2015}. However, none of these AF materials follow the genuine 2D Ising model {primarily due to either a small magnetic anisotropy or an XY-type anisotropy, along with interlayer interactions that impact the magnetic phase transition} \cite{Calder2021,Wildes2022}.

FePS$_3$ and FePSe$_3$ constitute a potential material family for exploring an Ising-type antiferromagnetic phase transition accompanied by $C_3$ symmetry breaking. This study specifically centers on FePSe$_3$[Figs. 1(a) and 1(c)]. FePSe$_3$ belongs to the rhombohedral $R\bar 3$ space group, with hexagonal layers stacked through weak vdW interactions along the $c$-axis [Fig. 1(b)] \cite{Wiedenmann1981}. Compared with FePS$_3$ in the monoclinic $C/2m$ symmetry group, FePSe$_3$ preserves the $C_3$ symmetry and therefore is ideal to study any possible 
magnetic order induced in-plane $C_3$ symmetry breaking without significant interlayer coupling effects \cite{Ouvrard1985}. Since FePSe$_3$ has an 
in-plane zig-zag AF structure below $T_N$=110K, we expect an FM exchange interaction $J_1$ between NN, and AF exchange interactions 
in $J_2$ and $J_3$ between NNN and NNNN, respectively [Fig. 1(a)] \cite{Wiedenmann1981}. In the 3D bulk limit, the AF ordering wavevector is ${\bf Q}_{AF}=(1/2,0,1/2)$ [Fig. 1(b) and 1(f)] due to 
a weak AF interlayer coupling $J_c$  [Fig. 1(b)] \cite{Wiedenmann1981}. The magnetic Fe$^{2+}$ ion has a 3$d^6$ electronic orbital with four unpaired spins, 
giving $S=2$. A strong spin-orbit coupling (SOC) induces a large magnetic anisotropy with 
the $c$-axis as the easy axis \cite{Wiedenmann1981,Bhutani2020}. Raman scattering experiments on bulk and monolayer FePSe$_3$ have observed a $\sim$15meV spin gap, suggesting that the system is Ising-like \cite{Luo2023}. 
 
In this work, we use elastic and inelastic neutron scattering (INS) to study the magnetic order and spin dynamics in bulk single crystal FePSe$_3$.
The temperature dependence of the magnetic order parameter from elastic neutron scattering experiments 
suggests that the AF phase transition is first-order in nature [Fig. 1(d)]. In the AF ordered state, our INS experiments reveal
that spin waves are gapped below $\sim$15meV, consistent with Raman scatteirng results \cite{Luo2023}, and are highly 2D with weak dispersion along the $c$-axis [Fig. 1(g)].  By fitting the spin-wave dispersion spectra using linear spin wave theory (LSWT) \cite{Holstein}, we determine magnetic exchange couplings, magnetic anisotropy, and find evidence for a biquadratic term in the spin Hamiltonian. 
On warming above $T_N$, we find dispersive spin excitations that can be explained by AF and FM excitations
from honeycomb lattice clusters with the $C_3$ symmetry. These results
are different from the expectation of a conventional AF second-order phase transition, 
 suggesting that the long-range AF order in FePSe$_3$ is replaced by $C_3$ magnetic honeycomb lattice clusters as the temperature is raised above $T_N$. The uncorrelated paramagnetic scattering from individual Fe ions
is only established at temperatures well above $T_N$.

Single crystals of FePSe$_3$ are synthesized using the chemical vapor transport method described in ref. \cite{Du2015}.  Elastic and INS experiments were respectively performed at the CORELLI spectrometer \cite{CORELLI} on one single crystal sample and the ARCS spectrometer\cite{ARCS} on $\sim$0.5g of co-aligned crystals at the Spallation Neutron Source, Oak Ridge National Laboratory. The momentum transfer $\boldsymbol{Q}$ is referenced in reciprocal lattice units (rlu) with respect to the rhombohedral unit cell of FePSe$_3$ with $a=b=$6.26\AA, $c=$19.71\AA. The magnetic ordering wavevector observed at ${\bf Q}_{AF}=(1/2,0,1/2)$ confirms the zig-zag magnetic structure, and the temperature dependence of the order parameter is shown in Fig. 1(d). A power law fit of the order parameter with $I=I_0(1-(T/T_N)^{2\beta})$ yields a critical exponent $\beta=0.063$, which is much smaller than the 2D Ising model prediction $\beta=1/8$, suggesting a first-order nature of the phase transition. This is further supported by measurements of correlation lengths which jump abruptly at $T_N$\cite{Bhutani2020}.

\begin{figure*}
\centering
\includegraphics[scale=.5]{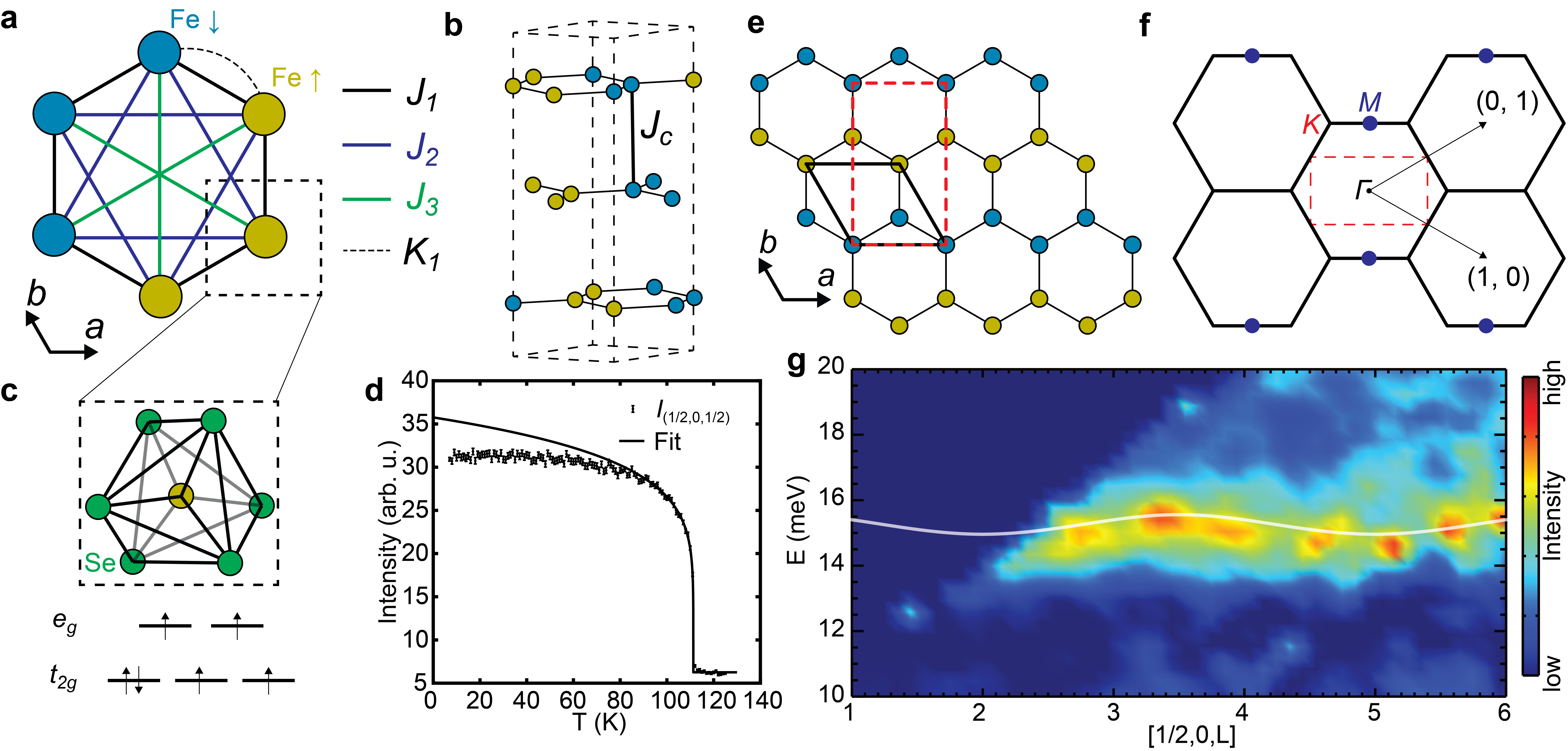}
\caption{\label{fig1} {\bf Real/reciprocal space of FePSe$_3$ and spin waves along in-plane and $c$-axis directions.} (a) Schematics of the intralayer AF magnetic structure and exchange interactions. (b) Interlayer magnetic structure and coupling $J_c$. (c) Local Fe$^{2+}$ environment in FePSe$_3$, with the 3$d^6$ spin configuration in an octahedral environment. (d) Neutron intensity of magnetic Bragg peak (1/2 0 1/2) as a function of temperature. {  (e) The in-plane zig-zag magnetic structure. The bold black rhombus shows the lattice unit cell and the dashed red rectangle shows the magnetic unit cell. (f) The reciprocal space with black hexagons showing the lattice FBZ and the dashed red rectangle shows the magnetic FBZ. The blue dots are the magnetic Bragg peaks from the magnetic structure shown in (e).} (g) Out-of-plane spin wave dispersion along [1/2 0 $L$] from the experiment.}
\end{figure*}

Figures 1(g) and 2(b) show the measured spin wave dispersions of FePSe$_3$ at 5 K along the $c$ axis and within the 2D honeycomb lattice plane, respectively. The overall dispersion has a band top of $\sim$40 meV, with an anisotropy gap of $\sim$15 meV at the $\Gamma$ and $M$ point. The large anisotropy gap value is also observed in the sister compound FePS$_3$, which is due to the combined effects of SOC in the 3d$^6$ orbital of Fe$^{2+}$ and the distortion of the FeSe$_6$ octahedron [Fig. 1(c)]. For comparison, the isostructural compounds MnPSe$_3$ and MnPS$_3$ do not exhibit such large gaps because their 3d$^5$ electrons have quenched orbital moments \cite{Calder2021,Wildes1998}. The spin waves propagating along the $c$-axis exhibit significantly less dispersion in contrast to the in-plane spin waves, featuring a bandwidth of approximately 1 meV for the former as opposed to a 20 meV bandwidth for the latter, indicating the presence of a very weak interlayer coupling $J_{c}$ [Fig. 1(g)]. This is consistent with the fact that the $T_N$ of FePSe$_3$ changes little as a function of layer numbers, suggesting that the interlayer exchange interactions have minimal effect on the magnetic phase transition in FePSe$_3$ \cite{Lee2016, Luo2023}.

To describe the in-plane spin wave dispersion, we first use LSWT to calculate the spin waves with the spin Hamiltonian

\begin{equation}
    H_0 = \sum_{<i,j>}J_{ij}\textbf{S}_{i}\cdot\textbf{S}_{j} + \sum_iD_{z}(S_{i}^{z})^{2}
\end{equation}
where the bilinear (Heisenberg) exchange interaction $J_{ij}$ is summed over the 1st, 2nd and 3rd NN, and $D_{z}$ is the single-ion anisotropy with $z$-axis as its easy axis. {   It is worth noting that within the actual system, magnetic anisotropy may stem from either single-ion effects or Ising-type exchanges. Therefore, the Hamiltonian can encompass Ising exchange terms $J_zS_{iz}S_{jz}$ in addition to the single-ion anisotropy. Nevertheless, within the framework of linear spin wave theory, the Ising term's impact on dispersion will be identical to that of single-ion anisotropy, provided that the respective parameters are appropriately adjusted (See the supplementary information). For the sake of simplicity, we have chosen to solely include the single-ion anisotropy term here.} Using a least-square-error fitting method with $S=2$, we extract the magnetic exchange coupling parameters as shown in TABLE I. However, the best fitting using this model with uniform $J_1$ does not precisely reproduce the dispersion, especially the low-energy part perpendicular to the zig-zag direction. In fact, the exact same scenario is observed in the sister compound FePS$_3$ \cite{Wildes2020} where a simple Heisenberg model plus anisotropy cannot account for the spin excitations accurately. In the case of FePS$_3$, two approaches have been utilized to resolve this problem, and here we apply them to FePSe$_3$ as well. The first is to introduce bond-dependent $J_1$ with $J_{1a}$ bonding the parallel spins and $J_{1b}$ for anti-parallel spins; The second is to introduce a biquadratic interaction 

\begin{equation}
    H = H_0 + \sum_{<i,j>} K_{ij}(\textbf{S}_{i}\cdot\textbf{S}_{j})^2
\end{equation}
where the biquadratic exchange $K_{ij}$ is summed over the 1st NN only. Both methods stabilize the zig-zag AF order, and with proper fitting parameters in TABLE I, they yield nearly identical dispersions that can accurately reproduce the experimental data. For the $J_{1a}$-$J_{1b}$ model, the difference between $J_{1a}$ and $J_{1b}$ is large, indicating that a lattice distortion should accompany the magnetic phase transition as in the case of FePS$_3$. However, no obvious lattice parameter change is observed across $T_N$, indicating that the $J_{1a}$-$J_{1b}$ model is unlikely to be correct. On the other hand, The existence of biquadratic interaction has been theoretically proposed in many such 2D magnetic systems with edge-shared octahedron structures\cite{Kartsev2020, Ni2021}, and therefore may be the more suitable model for describing the spin waves in FePSe$_3$. The situation is similar in the case of iron-based superconductors, where a biquadratic interaction has been used to account for the observed in-plane spin wave dispersions \cite{JZhao2009,CLiu2020}. {
A noteworthy fact is that the low-energy magnons at the $\Gamma$ point are linearly coupled to phonons, which introduces a magnon-polaronic gap that lifts the 2-fold degeneracy of the AF magnons\cite{Luo2023}. This suggests the necessity of introducing a magnon-phonon coupling term into the spin Hamiltonian. However, the impact of this term is minimal, exhibiting a gap of approximately 0.6 meV, a value below the resolution of the instruments. Hence, it is omitted from equation (2). Furthermore, the required strength of the Kitaev interaction (0.03 meV) to induce the magnon-phonon gap is significantly less than that of the Heisenberg exchanges, and thus, these are not taken into account in the fitting parameters.}

\begin{figure}
\centering
\includegraphics[scale=.43]{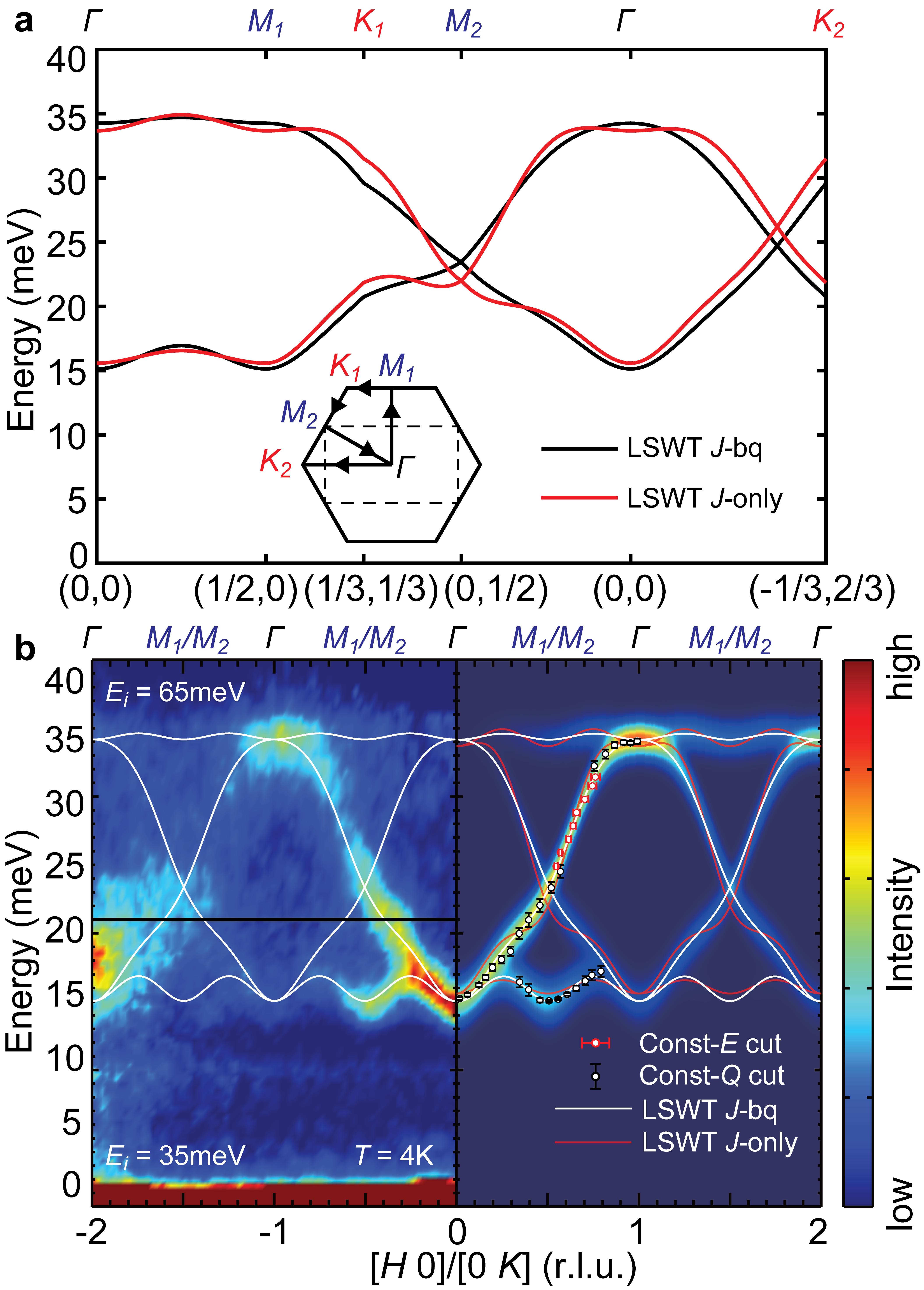}
\caption{{\bf Spin waves in FePSe$_3$.} (a) Calculated spin wave dispersion with (black) and without (red) biquadratic interactions. The inset shows the $\boldsymbol Q$ scan path. (b) LSWT fit of the experimental data with (white) and without (red) biquadratic interactions.}
\end{figure}


\begin{table}[t]
    \centering
    \begin{tabular}{|c|c|c|c|c|c|}
    \hline
       Model & $J_1$ & $J_2$ & $J_3$ & $K_1$ & $D_z$\\
    \hline
        Uniform $J_1$ & -2.30 & -0.23   & 2.01 & 0 &  -2.74\\
    \hline
        $J_{1a}$-$J_{1b}$& -2.26/-0.72 &  0.09  &  1.28 & 0 & -2.45\\
    \hline
        $J_1$-bq & -1.32  & 0.12 &   1.28    & -0.22 & -2.31\\
    \hline
    \end{tabular}
    \caption[Magnetic exchange interaction strength in different models in FePSe$_3$.]{Magnetic exchange interaction strength in different models in FePSe$_3$. All units in meV. In the $J_{1a}$-$J_{1b}$ model, the $J_{1a}$ indicates FM interactions in the zig-zag chain, and $J_{1b}$ indicates interactions between chains. {
    $J_{1}$, $J_{2}$, $J_{3}$ indicate the first, second, and third nearest neighbor Heisenberg exchange, respectively. $K_{1}$ refers to the nearest-neighbor biquadratic interactions, and $D_{z}$ stands for single-ion anisotropy.}}
    \label{tab:table3}
\end{table}



The first-order nature of the AF phase transition also displays itself in the spin fluctuations in the neighborhood of $T_N$. Fig. 3 shows a summary of the INS spectrum near $T_N = 110$ K. At $T=101$ K, in-plane spin excitations are well-defined and similar to spin waves at 5 K [Fig. 3(a)]. On warming to $T = 107$ K, spin excitations remain well defined, but become broader with an anisotropy spin gap above 10 meV [Fig. 3(b)]. 
Upon further warming to 110 K, the spin gap drastically decreases to zero at the $M$ point, while keeping a non-zero value at the $\Gamma$ point [Fig. 3(c)]. 
A cut along the $[H,0]$ direction at $E=5\pm 1$ meV shows clearly that the $M$ point spin fluctuations are enhanced from 110 K to 125 K, different
from the expectation of critical magnetic scattering associated with a second-order phase transition. Although the spin gap closing at the $M$ point may arise from the vanishing static magnetic moment across $T_N$, the differences in temperature dependence of the spin gap at $\Gamma$ and $M$ points cannot be explained by a second-order AF phase transition
since the $\Gamma$ and $M$ points are equivalent within the spin wave theory which requires long-range AF order and folding of the Brillouin zone below $T_N$. This is consistent with the observed first-order transition in the AF order parameter(fig.1d). In the paramagnetic state, the rectangular magnetic Brillouin zone unfolds to the hexagonal lattice Brillouin zone. However, we find similar spin excitations as in the AF ordered state [Figs. 3(d), 3(e), and 3(f)].

\begin{figure}
\centering
\includegraphics[scale=.32]{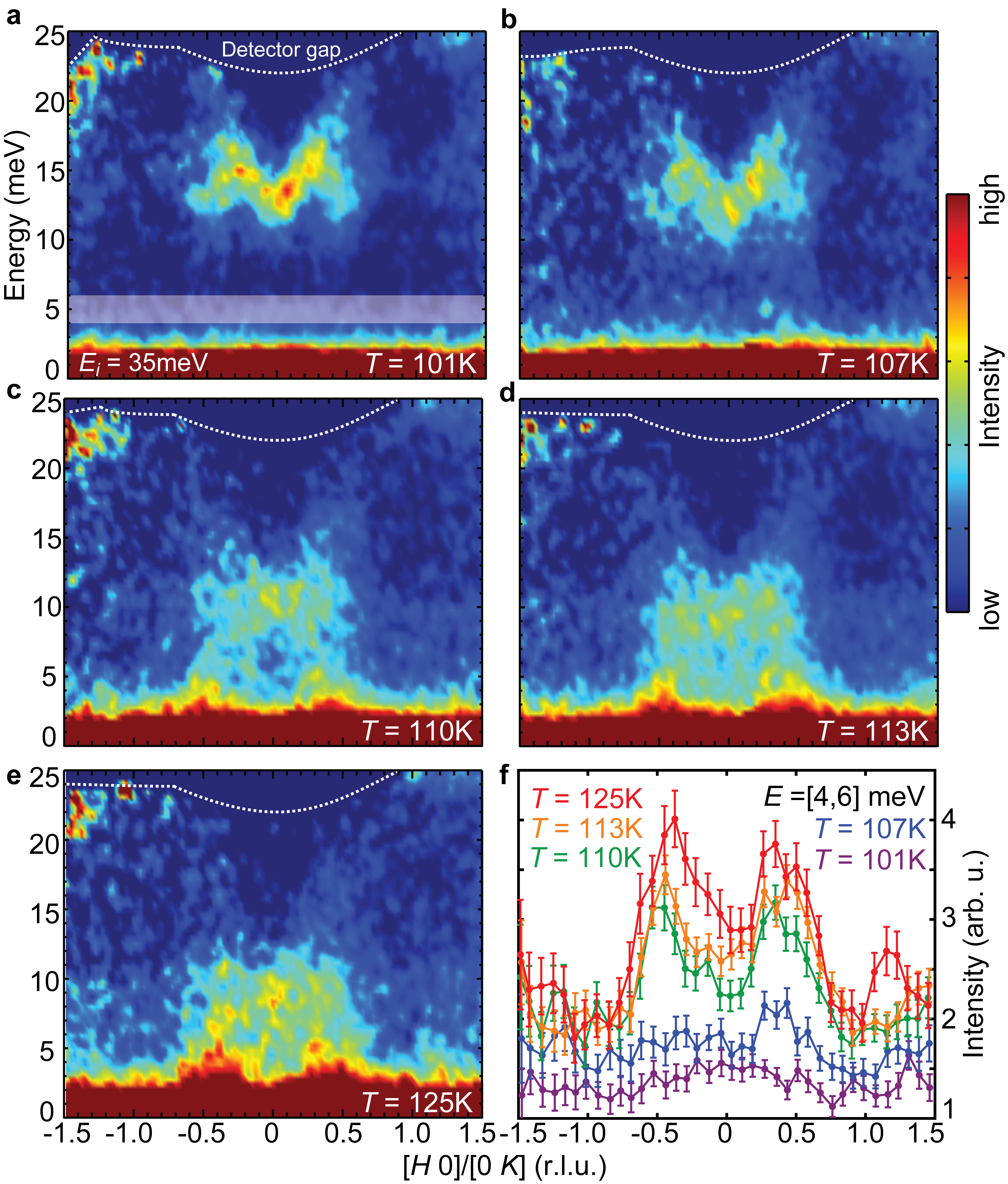}
\caption{{\bf Spin excitations in FePSe$_3$ in different temperatures.}Spin excitations in FePSe$_3$ in (a) 101K, (b) 107K,(c) 110K, (d) 113K, and (e) 125K. (f) a constant-$E$ cut of the intensity in (a-e) with $E$ integrated between 4-6meV. The incident neutron energy used here is $E_i$=35meV.
  }
\end{figure}
Although the AF phase transition in FePSe$_3$ is first-order, the short-range zig-zag order may still persist near the phase transition.
A natural way to reconcile the existence of the zig-zag order and the $C_3$ symmetry is to combine all the zig-zag domains with equal weight [Fig. 4(a)], while the size of the fluctuation domains determines the correlation lengths of the short-range order. 
To understand the nature of the dispersive paramagnetic spin excitations in FePSe$_3$, we plot constant-$E$ slices in the [$H$ $K$] plane at $E = 5\pm2$ meV [Fig. 4(c)] and $11\pm 2$ meV [Fig. 4(d)] at 110 K, which shows the spin excitations at the $M$ points and their connections, respectively.
The low-energy spin excitations show diffusive hexagonal patterns which are hollow at the $\Gamma$ point, and the high-energy excitations show plate-like hexagonal patterns. Inspired by the spin cluster methods used in refs. \cite{Lee2002,Yao2022,Gao2023}, we apply a similar analysis on FePSe$_3$. Assuming that the spins form clusters of hexagons with zig-zag order [Fig. 4(a), we calculate the neutron intensity $I(\boldsymbol{Q})$ with
\begin{equation}
    I(\boldsymbol Q) \propto f^2(\boldsymbol Q)\sum_{\langle m,n\rangle}e^{i\boldsymbol Q(\boldsymbol r_m-\boldsymbol r_n)}(1 - \frac{Q^2_z}{Q^2})\langle S_m^z S_n^z\rangle 
\end{equation}
where $f^2(\boldsymbol Q)$ is the magnetic form factor, $m,n\in{1,...,6}$ is the index for spins in Figs. 4(a) and 4(b). The calculated $I(\boldsymbol Q)$ is then averaged over six zig-zag spin configurations. Figures 4(d) and 4(f) show the calculated intensity from the zig-zag clusters [Fig. 4(a)], which matches well with the experimental result at low energy. For the higher energy part at 11meV, due to the enhanced $\Gamma$ point spin fluctuations, an additional FM cluster is required [Fig. 4(b)] in order to reconstruct the plate-like excitation pattern.       
\begin{figure}[t]
\centering
\includegraphics[scale=.35]{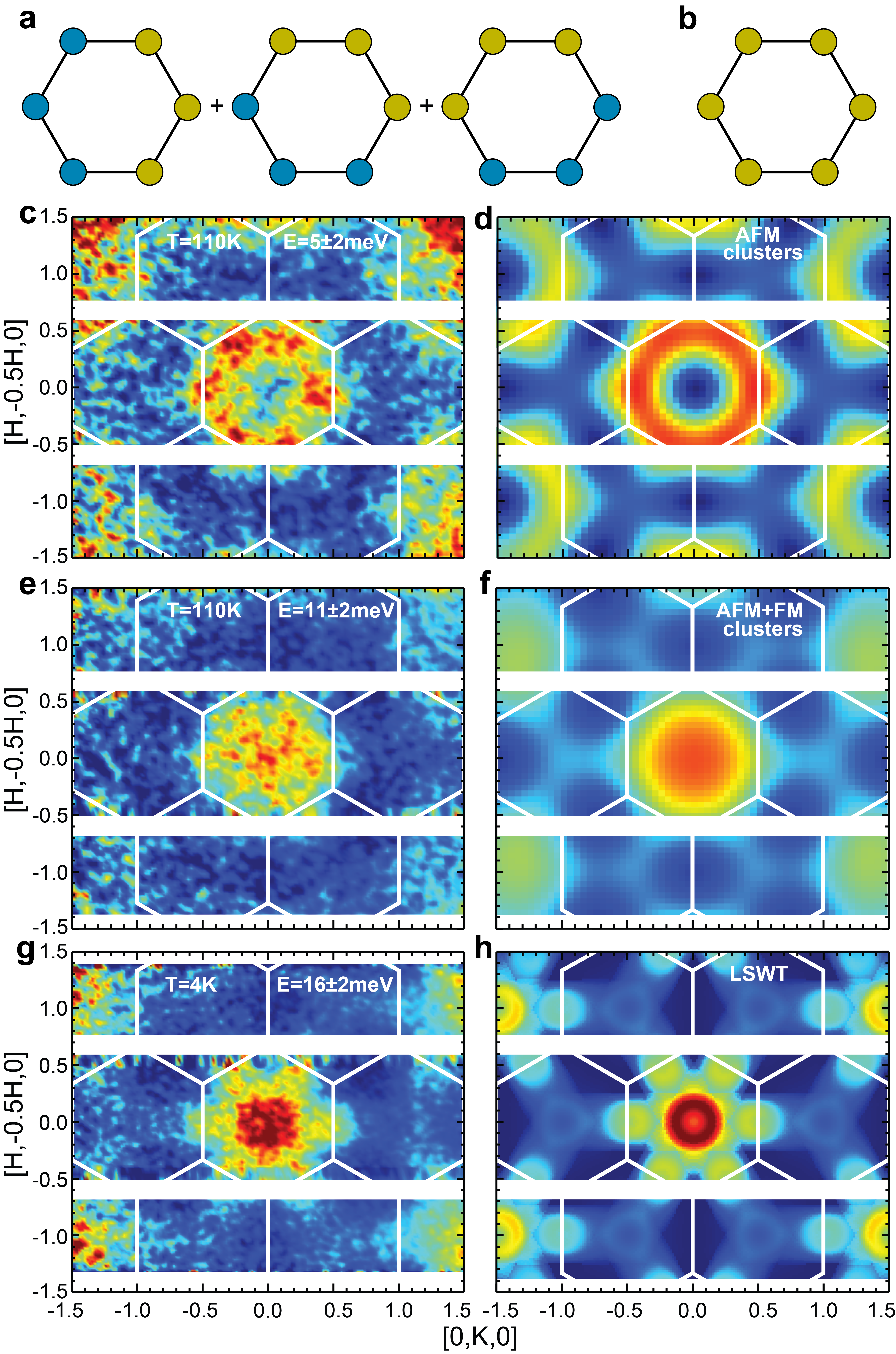}
\caption{{\bf Simulation on spin fluctuations at Neel temperature in FePSe$_3$.} (a,b) Zig-zag AF and FM clusters used to simulate the spin fluctuation, respectively. (c) Experimental data at $T$=110K and $E$=[3,7]meV. (d) Calculated neutron intensity of the AF cluster in (a). (e) Experimental data at $T$=110K and $E$=[9,13]meV. (f) Calculated neutron intensity of the AF cluster (a) plus the FM cluster (b). The AF clusters and FM clusters are weighted with a ratio of 2:1 in (f). {  (g) experimentalw3 data at $T$=4K and $E$=[14,18]meV. (h) Calculated neutron intensity at the same energy range as (g) by LSWT with parameters from TABLE I.}}
\end{figure}

{The spin fluctuations observed at 110K encode substantial information regarding the spin correlations in FePSe$_3$. The pillar-like feature of the spin excitations at the $M$ point in Figure 3(c) indicates that the $\boldsymbol{Q}$-dependence of low-energy spin excitations remains unchanged below $\sim$8meV. This implies that an interval over the low-energy excitations can reasonably approximate the instantaneous spin correlations. Consequently, despite our zig-zag cluster calculation being used to reproduce the spin fluctuation in the energy range $E$=[3,7] meV, it is reasonable to infer that this zig-zag arrangement accurately reflects the nature of these instantaneous spin correlations.

A noteworthy observation is that even in the AF state at 4K, the dynamical spin-spin correlation centers around the FM wavevector located at $\Gamma$, rather than the AF wavevector at $M$ [Figure 4(g) and 4(h)]. The reason is that as the magnetic anisotropy grows, the AF spin wave excitations are gradually replaced by local spin-flip excitations. In an isotropic AF Heisenberg model, the magnetic excitations are spin waves that give zero structure factor at the FM wavevector, while in a pure AF Ising model, the spin-flip excitations will introduce a non-zero intensity centered at the $\Gamma$ point. In the case of FePSe$_3$, the robust magnetic anisotropy shifts the primary spin excitations from AF spin waves to spin-flip excitations. As the temperature rises to $T_N$, the local spin-flip excitation remains gapped, indicating that the dynamic susceptibility associated with spin-flip scattering remains non-zero. Consequently, it cannot be the primary driving force behind the first-order magnetic phase transition. Instead, the excitations of zig-zag clusters at $T_N$ become gapless, implying that the dominant component of spin configurations near $T_N$ is the zig-zag short-range order. It can be speculated that the magnetic phase transition is driven by a discontinuous condensation of these zig-zag hexagonal clusters, corresponding to the $C_3$-$C_1$ symmetry breaking. This distinguishes it from the fluctuations observed in conventional Ising magnets that break the $Z_2$ time-reversal symmetry through spin-flip scatterings.}

In summary, we utilized neutron scattering to study the spin excitations and spin fluctuations in FePSe$_3$ across $T_N$. By analyzing the spin wave dispersion, we find evidence for biquadratic magnetic exchange interactions in the effective spin Hamiltonian terms. The microscopic origin of the biquadratic term remains to be determined.
In addition, our magnetic order parameter measurements indicate that the AF phase transition is first-order in nature. From measurements of the temperature-dependent spin excitations across $T_N$, we infer that the AF phase transition may be driven by the $C_3$ to $C_1$ symmetry breaking from low-energy spin clusters associated with zig-zag AF honeycomb lattice, respectively. {  Our results provide an enriched perspective on the intricate interplay between magnetic interactions and structural symmetries in the Ising magnetic systems with additional $C_3$ symmetry breaking.}

The neutron scattering and single-crystal synthesis work at Rice was supported by US
NSF-DMR-2100741 (P.D.) and by the Robert A. Welch Foundation under
grant no. C-1839 (P.D.), respectively. {
A part of this research is supported by the Edinburgh-Rice Strategic Collaboration Award} (P.D.). The work at the University of
California, Berkeley was supported by the U.S. DOE under contract
no. DE-AC02-05-CH11231 within the Quantum Materials Program
(KC2202) (R.J.B.). X.T. and M.Y. are supported by the Gordon and Betty Moore Foundation's EPiQS Initiative through grant No. GBMF9470 and the Robert A. Welch Foundation Grant No. C-2175. The work of J.-H.C. was supported by the National Research Foundation (NRF) of Korea (Grant nos. 2020R1A5A1016518 and 2022R1F1A1074321). The research at Hangzhou Normal University is supported by The Open Project of Guangdong Provincial Key Laboratory of Magnetoelectric Physics and Devices, No. 2022B1212010008
, Startup Project of Hangzhou Normal University (Grant No. 2020QDL026) and Natural Science Foundation of Zhejiang Province (Grant No. LY22A040009). A portion of this research used resources at the Spallation Neutron
Source, a DOE Office of Science User Facility operated by Oak Ridge National Laboratory.

{}


\begin{thebibliography}{}
\bibitem{Onsager1944} Lars Onsager, Crystal Statistics. I. A Two-Dimensional Model with an Order-Disorder Transition \textit{Phys. Rev.} {\bf 65}, 117 (1944).

\bibitem{B} V. L. Berezinskii, Destruction of long-range order in one-dimensional and two-dimensional systems having a continuous symmetry group I. Classical systems,  Sov. Phys. JETP, \textbf{32} (3): 493–500 (1970).

\bibitem{KT}J. M. Kosterlitz and D. J. Thouless, Ordering, metastability and phase transitions in
two-dimensional systems. J. Phys. C: Solid State Phys., \textbf{6}, 1181-1203 (1973).

\bibitem{Gibertini2019} M. Gibertini, M. Koperski, A. F. Morpurgo, and K. S. Novoselov, Nat. Nano. {\bf 14}, 408 (2019).

\bibitem{Mermin} N. D. Mermin and H. Wagner, Absence of Ferromagnetism or Antiferromagnetism in One- or Two-Dimensional Isotropic Heisenberg Models. \textit{Phys. Rev. Lett.} \textbf{17}, 1133 (1966).

\bibitem{Goldenfeld1972} Nigel Goldenfeld, \textit{Lectures On Phase Transitions And The Renormalization Group}, CRC Press (1972).

\bibitem{Huang2017}B. Huang, G. Clark, E. Navarro-Moratalla, D. R. Klein, R. Cheng, K. L. Seyler, D. Zhong, E. Schmidgall, M. A. McGuire, D.H. Cobden, W. Yao, D. Xiao, P. Jarillo-Herrero, and X. Xu, Layer-Dependent Ferromagnetism in a Van der Waals Crystal Down to the Monolayer Limit, \textit{Nature} \textbf{546}, 270 (2017).

\bibitem{Gong2017} C. Gong, L. Li, Z. Li, H. Ji, A. Stern, Y. Xia, T. Cao, W. Bao, C. Wang, Y. Wang, Z. Q. Qiu, R. J. Cava, S. G. Louie, J. Xia, and X. Zhang, Discovery of Intrinsic Ferromagnetism in Two-Dimensional Van der Waals Crystals, \textit{Nature} \textbf{546}, 265 (2017).

\bibitem{Lee2016} Jae-Ung Lee, Sungmin Lee, Ji Hoon Ryoo, Soonmin Kang, Tae Yun Kim, Pilkwang Kim, Cheol-Hwan Park, Je-Geun Park, and Hyeonsik Cheong, Ising-Type Magnetic Ordering in Atomically Thin FePS$_3$. \textit{Nano Lett.} \textbf{16}, 12, 7433–7438 (2016).

\bibitem{Lee2023} Youjin Lee, Suhan Son, Chaebin Kim, Soonmin Kang, Junying Shen, Michel Kenzelmann, Bernard Delley, Tatiana Savchenko, Sergii Parchenko, Woongki Na, Ki-Young Choi, Wondong Kim, Hyeonsik Cheong, Peter M. Derlet, Armin Kleibert, Je-Geun Park, Giant Magnetic Anisotropy in the Atomically Thin van der Waals Antiferromagnet FePS$_3$. \textit{Adv. Electron. Mater.} \textbf{9}, 2200650 (2023).

\bibitem{Luo2023}Jiaming Luo, Shuyi Li, Zhipeng Ye, Rui Xu, Han Yan, Junjie Zhang, Gaihua Ye, Lebing Chen, Ding Hu, Xiaokun Teng, William A. Smith, Boris I. Yakobson, Pengcheng Dai, Andriy H. Nevidomskyy, Rui He, and Hanyu Zhu. Evidence for Topological Magnon–Phonon Hybridization in a 2D Antiferromagnet down to the Monolayer Limit. \textit{Nano Lett}. \textbf{23}, 5, 2023–2030 (2023).

\bibitem{Fouet2001} J. B. Fouet, P. Sindzingre, C. Lhuillier. An investigation of the quantum $J_1$-$J_2$-$J_3$ model on the honeycomb lattice. \textit{ Eur. Phys. J. B } \textbf{20}, 241–254 (2001).

\bibitem{Domany1978} E. Domany, E. K. Riedel, Phase transitions in two‐dimensional systems. \textit{Journal of Applied Physics }\textbf{49}, 1315–1320 (1978).

\bibitem{Liu2016} Bingjie Liu, Youming Zou, Lei Zhang, Shiming Zhou, Zhe Wang, Weike Wang, Zhe Qu and Yuheng Zhang, Critical behavior of the quasi-two-dimensional semiconducting ferromagnet CrSiTe$_3$. \textit{Sci. Rep.} \textbf{6}, 33873 (2016). 

\bibitem{Williams2015} T. J. Williams, A. A. Aczel, M. D. Lumsden, S. E. Nagler, M. B. Stone, J.-Q. Yan, and D. Mandrus, Magnetic correlations in the quasi-two-dimensional semiconducting ferromagnet 
CrSiTe$_3$. \textit{Phys. Rev. B} \textbf{92}, 144404 (2015).

\bibitem{Ron2019}Ron, A., Zoghlin, E., Balents, L.\textit{ et al.}, Dimensional crossover in a layered ferromagnet detected by spin correlation driven distortions. \textit{Nat Commun} \textbf{10}, 1654 (2019).

\bibitem{Liu2018} Yu Liu and C. Petrovic, Three-dimensional magnetic critical behavior in 
CrI$_3$ \textit{Phys. Rev. B} \textbf{97}, 014420 (2018).

\bibitem{Chen2020}Lebing Chen, Jae-Ho Chung, Tong Chen, Chunruo Duan, Astrid Schneidewind, Igor Radelytskyi, David J. Voneshen, Russell A. Ewings, Matthew B. Stone, Alexander I. Kolesnikov, Barry Winn, Songxue Chi, R. A. Mole, D. H. Yu, Bin Gao, and Pengcheng Dai, Magnetic anisotropy in ferromagnetic CrI$_3$. \textit{Phys. Rev. B} \textbf{101}, 134418 (2020).

\bibitem{Liu2020} Yu Liu, Milinda Abeykoon, and C. Petrovic Critical behavior and magnetocaloric effect in VI$_3$. \textit{Phys. Rev. Research} \textbf{2}, 013013 (2020).

\bibitem{Hao2021} Yiqing Hao \textit{et al.}, Magnetic Order and Its Interplay with Structure Phase Transition in van der Waals Ferromagnet VI$_3$. \textit{Chinese Phys. Lett}. \textbf{38} 096101 (2021).

\bibitem{Lin2017} G. T. Lin, H. L. Zhuang, X. Luo, B. J. Liu, F. C. Chen, J. Yan, Y. Sun, J. Zhou, W. J. Lu, P. Tong, Z. G. Sheng, Z. Qu, W. H. Song, X. B. Zhu, and Y. P. Sun, Tricritical behavior of the two-dimensional intrinsically ferromagnetic semiconductor CrGeTe$_3$. \textit{Phys. Rev. B} \textbf{95}, 245212 (2017).

\bibitem{Wildes2007} A.R. Wildes \textit{et al.} Anisotropy and the critical behaviour of the quasi-2D antiferromagnet, MnPS$_3$. \textit{Journal of Magnetism and Magnetic Materials} \textbf{310} 1221–1223 (2007).

\bibitem{Wildes2017} A. R. Wildes, V. Simonet, E. Ressouche, R. Ballou and G. J. McIntyre, The magnetic properties and structure of the quasi-two-dimensional antiferromagnet CoPS$_3$. \textit{J. Phys.: Condens. Matter} \textbf{29} 455801 (2017).

\bibitem{Wildes2015} A. R. Wildes, V. Simonet, E. Ressouche, G. J. McIntyre, M. Avdeev, E. Suard, S. A. J. Kimber, D. Lançon, G. Pepe, B. Moubaraki, and T. J. Hicks, Magnetic structure of the quasi-two-dimensional antiferromagnet NiPS$_3$. \textit{Phys. Rev. B} \textbf{92}, 224408 (2015).

\bibitem{Calder2021} S. Calder, A. V. Haglund, A. I. Kolesnikov, and D. Mandrus. Magnetic exchange interactions in the van der Waals layered antiferromagnet MnPSe$_3$. Phys. Rev. B, 103:024414, (2021).

\bibitem{Wildes2022} A. R. Wildes, J. R. Stewart, M. D. Le, R. A. Ewings, K. C. Rule, G. Deng, and K. Anand, Magnetic dynamics of NiPS$_3$. \textit{Phys. Rev. B} \textbf{106}, 174422 (2022).

\bibitem{Wiedenmann1981} A. Wiedenmann, J. Rossat-Mignod, A. Louisy, R. Brec, J. Rouxel, Neutron diffraction study of the layered compounds MnPSe$_3$ and FePSe$_3$. \textit{Solid State Commun.} \textbf{40}, 1067– 1072 (1981).

\bibitem{Ouvrard1985} G. Ouvrard, R. Brec and J. Rouxel, Structural determination of some MPS$_3$ layered phases (M = Mn, Fe, Co, Ni and Cd). \textit{Mat. Res. Bull.}, \textbf{20}, 1181-1189, (1985).

\bibitem{Bhutani2020} Ankita Bhutani, Julia L. Zuo, Rebecca D. McAuliffe, Clarina R. de la Cruz, and Daniel P. Shoemaker, Strong anisotropy in the mixed antiferromagnetic system Mn$_{1-x}$Fe$_{x}$PSe$_{3}$, \textit{Phys. Rev. Materials} \textbf{4}, 034411 (2020).

\bibitem{Holstein} Holstein, T. \&  Primakoff, H. Field dependence of the intrinsic domain magnetization of a ferromagnet. \textit{Phys. Rev.} {\bf 58}, 1098 (1940). 

\bibitem{Du2015} Kezhao Du, Xingzhi Wang, Yang Liu, Peng Hu, M. Iqbal Bakti Utama, Chee Kwan Gan, Qihua Xiong, and Christian Kloc. Weak van der Waals stacking, wide-range band gap, and Raman study on ultrathin layers of metal phosphorus trichalcogenides. \textit{ACS Nano}, \textbf{10}(2):1738–1743, (2015).

\bibitem{CORELLI} Feng Ye, Yaohua Liu, Ross Whitfield, Ray Osbornb and Stephan Rosenkranz, Implementation of cross correlation for energy discrimination on the time-of-flight spectrometer CORELLI. \textit{J. Appl. Cryst.} \textbf{51}, 315-322 (2018).

\bibitem{ARCS} D. L. Abernathy; M. B. Stone; M. J. Loguillo; M. S. Lucas; O. Delaire; X. Tang; J. Y. Y. Lin; B. Fultz, Design and operation of the wide angular-range chopper spectrometer ARCS at the Spallation Neutron Source. \textit{Rev Sci Instrum.} \textbf{83}, 015114 (2012).


\bibitem{Wildes1998}A R Wildes, B Roessli, B Lebech, and K W Godfrey. Spin waves and the critical behaviour of the magnetization in MnPS$_3$. \textit{Journal of Physics: Condensed Matter}, \textbf{10}(28):6417–6428, (1998).

\bibitem{Leonel2006} S.A. Leonel, Amanda Castro Oliveira, B.V. Costa, Pablo Zimmermann Coura, Comparative study between a two-dimensional anisotropic Heisenberg antiferromagnet with easy-axis single-ion anisotropy and one with easy-axis exchange anisotropy. \textit{Journal of Magnetism and Magnetic Materials} \textbf{305} 157–164 (2006). 

\bibitem{Wildes2020} A. R. Wildes, M. E. Zhitomirsky, T. Ziman, D. Lan¸con, and H. C. Walker. Evidence for biquadratic exchange in the quasi-two-dimensional antiferromagnet
FePS$_3$. \textit{Journal of Applied Physics}, \textbf{127}(22):223903, (2020).

\bibitem{JZhao2009} Jun Zhao, DT Adroja, Dao-Xin Yao, R Bewley, Shiliang Li, XF Wang, G Wu, XH Chen, Jiangping Hu, Pengcheng Dai, Spin waves and magnetic exchange interactions in CaFe$_2$As$_2$, Nat. Phys. {\bf 5}, 555 (2009).

\bibitem{CLiu2020} Changle Liu, Xingye Lu, Pengcheng Dai, Rong Yu, and Qimiao Si, Anisotropic magnetic excitations of a frustrated bilinear-biquadratic spin model: Implications for spin waves of detwinned iron pnictides, Phys. Rev. B {\bf 101}, 024510 (2020).

\bibitem{Kartsev2020} Alexey Kartsev, Mathias Augustin, Richard F. L. Evans, Kostya S. Novoselov and Elton J. G. Santos, Biquadratic exchange interactions in two-dimensional magnets. \textit{npj Computational Materials} \textbf{6}, 150 (2020).

\bibitem{Ni2021} J. Y. Ni, X. Y. Li , D. Amoroso , X. He, J. S. Feng, E. J. Kan, S. Picozzi, and H. J. Xiang, Biquadratic exchange interactions in two-dimensional magnets. \textit{Phys. Rev. Lett.} \textbf{127}, 247204 (2021).

\bibitem{Straley1973} J. P. Straley and Michael E. Fisher, Three-state Potts model and anomalous tricritical points.\textit{ J. Phys. A: Math., Nucl. Gen.}, \textbf{6}, 1310-1326 (1973).

\bibitem{Domany1977} Eytan Domany, M. Schick, and J. S. Walker, Classification of Order-Disorder Transitions in Common Adsorbed Systems: Realization of the Four-State Potts Model.  \textit{Phys. Rev. Lett}. \textbf{38}, 1148 (1977).

\bibitem{Banerjee1964} Banerjee, B. K. On a generalised approach to first and second order magnetic transitions. \textit{Physics Letters}, \textbf{12}(1), 16–17. (1964).

\bibitem{Wilson2009} Stephen D. Wilson, Z. Yamani, C. R. Rotundu, B. Freelon, E. Bourret-Courchesne, and R. J. Birgeneau, Neutron diffraction study of the magnetic and structural phase transitions in BaFe$_2$As$_2$. \textit{Phys. Rev. B} \textbf{79}, 184519 (2009).

\bibitem{Lee2002} {
S.-H. Lee, C. Broholm, W. Ratcliff, G. Gasparovic, Q. Huang, T. H. Kim and S.-W. Cheong, Emergent excitations in a geometrically frustrated magnet. \textit{Nature}\textbf{ 418}, 856–858 (2002).}

\bibitem{Yao2022} Weiliang Yao, Kazuki Iida, Kazuya Kamazawa, and Yuan Li, Excitations in the Ordered and Paramagnetic States of Honeycomb Magnet Na$_2$Co$_2$TeO$_6$. \textit{Phys. Rev. Lett.} \textbf{129}, 147202 (2022).

\bibitem{Gao2023} Bin Gao, Tong Chen, Xiao-Chuan Wu, Michael Flynn, Chunruo Duan, Lebing Chen, Chien-Lung Huang, Jesse Liebman, Shuyi Li, Feng Ye, Matthew B. Stone, Andrey Podlesnyak, Douglas L. Abernathy, Devashibhai T. Adroja, Manh Duc Le, Qingzhen Huang, Andriy H. Nevidomskyy, Emilia Morosan, Leon Balents, and Pengcheng Dai, Diffusive excitonic bands from frustrated triangular sublattice in a singlet-ground-state system, Nat. Comm. {\bf 14}, 2051 (2023).














\end{thebibliography}
\end{document}